\documentclass[prd,aps,showpacs,floats,floatfix,superscriptaddress,nofootinbib,eqsecnum]{revtex4}

\usepackage{graphicx}
\usepackage{amsmath}
\usepackage{amssymb}
\usepackage{amsfonts}
\usepackage{bbm}
\usepackage{amscd}
\usepackage{array}
\usepackage{tensind}
\usepackage{mathrsfs}
\tensordelimiter{?}

\usepackage{longtable}

\def\laq{\raise 0.4ex\hbox{$<$}\kern -0.8em\lower 0.62ex\hbox{$\sim$}}
\def\gaq{\raise 0.4ex\hbox{$>$}\kern -0.7em\lower 0.62ex\hbox{$\sim$}}

\newcommand{\ba}{\begin{array}}
\newcommand{\ea}{\end{array}}

\newcommand{\be}{\begin{equation}}
\newcommand{\ee}{\end{equation}}
\newcommand{\bea}{\begin{eqnarray}}
\newcommand{\eea}{\end{eqnarray}} 
\def\nl{\\ & \quad}

\def\vct#1{{\mathchoice{\mbox{\boldmath$#1$}}{\mbox{\boldmath$#1$}}%
  {\mbox{\scriptsize\boldmath$#1$}}{\mbox{\scriptsize\boldmath$#1$}}}}

\newcommand{\mytextrm}[1]{{}}

\def\ee{{\mathrm{e}}}

\def\dd{{\mathrm{d}}}

\newlength{\sizeonefig}
\newlength{\sizetwofig}
\newlength{\sizeonefigb}
\newlength{\sizetwofigb}
\begin{document}

\title{Higher-order-in-spin interaction Hamiltonians for binary black
holes from Poincar\'e invariance}

\author{Steven Hergt} 

\author{Gerhard Sch\"afer} 

\affiliation{Friedrich-Schiller-Universit\"at Jena, Max-Wien-Platz 1, 07743 Jena, Germany}

\begin{abstract}
The fulfillment of the space-asymptotic Poincar\'e algebra
is used to derive new higher-order-in-spin interaction Hamiltonians for
binary black holes in the Arnowitt-Deser-Misner canonical formalism almost completing the set of the formally $1/c^4$
spin-interaction Hamiltonians involving nonlinear spin terms. To linear
order in $G$, the expressions for the $S^3p$ and
the $S^2p^2$ Hamiltonians are completed. It is also shown
that there are no quartic nonlinear $S^4$ Hamiltonians to linear order in $G$.
\end{abstract}


\date{\today}

\maketitle

\section{Introduction}

In order to obtain higher accuracy in the templates for analyzing
gravitational waves from binary black holes, gravitational spin-interaction terms have to be taken into account beyond
the leading order ones which are of the formal order $1/c^2$ (in this counting, spins are treated of the order
zero in terms of $1/c^2$), where $c$ denotes the speed of light. At the formal
order of $1/c^4$, several Hamiltonians or Lagrangians have been determined
already: The spin-orbit coupling ones, $H_{SO} = H_{Sp} +
H_{Sp^3}$, are given in \cite{DJS}, also see \cite{FBB}, the
spin(1)-spin(2) coupling ones, $H_{S_1S_2} + H_{S_1S_2p^2}$, can be found in \cite{SSH}, also see \cite{PR1},  
and the following Hamiltonians $H_{S_2^2S_1p_1} + H_{S_2^3p_1} + H_{S_1^3p_2} + H_{S_1^2S_2p_2},~ H_{S_1^2S_2^2},~
H_{S_1S_2^3}+H_{S_2S_1^3}$ have been derived in \cite{HS}; recently,
the dynamics corresponding to the Hamiltonians $H_{S_1^2}+ H_{S_2^2} + H_{S_1^2p^2} + H_{S_2^2p^2}$
has been announced, \cite{PR2}.
In this paper the missing $S^3p$-Hamiltonians $H_{S_1^3p_1} +
H_{S_2^3p_2}$ and $H_{S_1^2S_2p_1} + H_{S_2^2S_1p_2}$
are calculated in the Arnowitt-Deser-Misner (ADM) canonical formalism by applying the space-asymptotic Poincar\'e
algebra, \cite{ADM}, \cite{RT}. These expressions contain post-Newtonian leading order quadrupole
deformation effects represented by spin-squared terms (the coefficients
in the quadrupole terms tell that the treated bodies are black holes, see, e.g. \cite{HS}, \cite{PR2}).
Because of the power of the Poincar\'e algebra in controlling the higher post-Newtonian
dynamics we are also able to give reasonable expressions for the
spin-squared $S^2p^2$-Hamiltonians to linear order in $G$, i.e. $H_{S_1^2p^2} + H_{S_2^2p^2}$.
Unfortunately, only after completion of the calculation of all static $G^2$-terms
$H_{S^2}$ a comparison with the result presented in \cite{PR2} can be
made because of quite different approaches.  

Our units are very often $c=1$ and also $16\pi G=1$, where $G$ is the Newtonian
gravitational constant. Greek indices will run over $0,1,2,3$, Latin
from the beginning of the alphabet over $1,2$ (denoting black-hole 
numbers) and from its midst over $1,2,3$. For the signature of spacetime
we choose +2. $r \equiv r_{ab} = |x^i_a-x^i_b|$ will denote the Euclidean distance
between black hole $a$ and $b$, and $r~n^i \equiv r_{ab} ~n^i_{ab} =
x^i_a-x^i_b, (n^i_{ab}) = \vct{n}_{ab} \equiv  \vct{n}$.


\section{The Poincar\'e invariance}

The starting point for our approach will be the Poincar\'e algebra. The
Hamiltonians we wish to calculate have to fulfill this algebra using
standard Poisson brackets for the fundamenal position, linear momentum
and spin variables, $x^i_a, p_{ai}, (p_{ai})= \vct{p}$, and $S_{ai}, (S_{ai}) = \vct{S}$, respectively, so the  
plan is to make the most general ansatz for these Hamiltonians and plug them into the algebra to see how restrictive it is
upon them. Of course, the Hamiltonians cannot be determined uniquely
because all Hamiltonians which are canonically equivalent 
fulfill the algebra, so there will be degrees of freedom left which can
be fixed only by choosing an appropriate representation. We will choose to work within the
ADM canonical formalism using generalized isotropic coordinates, \cite{ADM}. With
the aid of a reasonable ansatz for the source terms in the constraint
equations we are able to fix all the coefficients of the
Hamiltonians in question.

The Poisson bracket relations defining the Poincar\'e algebra read, see,
e.g. \cite{DJS00},

\begin{equation}
 \left\{P_{i},P_{j}\right\}=0,\qquad \left\{P_{i},H\right\}=0,\qquad \left\{J_{i},H\right\}=0,
\end{equation}

\begin{equation}
 \left\{J_{i},P_{j}\right\}=\epsilon_{ijk}P_{k}, \qquad \left\{J_{i},J_{j}\right\}=\epsilon_{ijk}J_{k},
\end{equation}

\begin{equation}
 \{J_{i},G_{j}\}=\epsilon_{ijk}G_{k},
\end{equation}

\begin{equation}
 \{G_{i},H\}=P_{i} \label{crucial},
\end{equation}

\begin{equation}
 \{G_{i},P_{j}\}=\frac{1}{c^2}H\delta_{ij},
\end{equation}

\begin{equation}
 \{G_{i},G_{j}\}=-\frac{1}{c^2}\epsilon_{ijk}J_{k}.
\end{equation}
The total linear momentum $P_i$ and the total angular momentum $J_{i}$
are respectively given by $P_i=\sum_a p_{ai}$ and  $J_i = \sum_a
J_{ai}$, where $J_{ai}  = \epsilon_{ijk}x^j_a p_{ak} + S_{ai}$. 
The crucial equation which needs to be checked and which finally will be used for
determining the new Hamiltonians is Eq. (\ref{crucial}). It describes the
time evolution of the center-of-mass vector $G_i$.
In the post-Newtonian context, the importance of Eq. (\ref{crucial}) for the
fixation of the Hamiltonian has been stressed already in Ref. \cite{JS00},
which is the source paper for \cite{DJS00}. Obviously,
the Hamiltonian in Eq. (2.5) needs to be known to one order in $1/c^2$ less
compared with Eq. (\ref{crucial}), and Eq. (2.1) only tells that the Hamiltonian has to be invariant
against 3-dimensional translations and rotations. The second post-Newtonian Hamiltonian (2PN-Hamiltonian)
of formal order $1/c^4$ which enters the equations above reads

\begin{equation}
 H=H_{N}+H_{1PN}+H_{2PN}+H_{SO}^{1PN}+H_{SO}^{2PN}+H_{S^2}+H_{S^3p}+H_{S^2p^2}+H_{S^4}
\end{equation}
with

\begin{equation}
 H_{S^2}=H_{S_{1}S_{2}}+H_{S_{1}^2}+H_{S_{2}^2}.
\end{equation}
The Hamiltonians $H_{N}$, $H_{1PN}$ and $H_{2PN}$ are the point particle
ones given in e.g. \cite{DJS00}. The Hamiltonians $H_{SO}^{1PN}$,
$H_{S_{1}S_{2}}$ and $H_{S_{1}^2}$ + $H_{S_{2}^2}$, as far as being of order 1PN, were
(re-)calculated in \cite{HS}, also see \cite{KWW}, \cite{D}. The
Hamiltonian $H_{SO}^{2PN}$ is known from \cite{DJS} and $H_{S_{1}S_{2}(p^2+G)}$ has been derived in
\cite{SSH}. The Hamiltonian $H_{S^3p}$ splits into the parts

\begin{equation}
 H_{S^3p}=H_{S_{1}^3p_1}+H_{S_{2}^3p_2}+H_{S_{2}^3p_1}+H_{S_{1}^3p_2}+H_{S_{1}S_{2}^2p_1}+H_{S_{1}^2S_{2}p_1}
 +H_{S_{2}S_{1}^2p_2}+H_{S_{2}^2S_{1}p_2}\,.
\end{equation}
and $H_{S^2p^2}$ is an abbreviation for

\begin{equation}
 H_{S^2p^2}=H_{S_{1}S_{2}p^2}+H_{S_{1}^2p^2}+H_{S_{2}^2p^2}.
\end{equation}
The contributions $H_{S_{2}^3p_1}$, $H_{S_{1}^3p_2}$,
$H_{S_{2}S_{1}^2p_2}$ and $H_{S_{1}S_{2}^2p_1}$ were all
calculated in \cite{HS} (${\vct{S}}_b = m_b {\vct{a}}_b$), 

\begin{equation}
H_{S_{1}^3p_2}=\;-G\vct{S}_{1}\!\cdot\!\left(\vct{n}_{12}\times\vct{p}_{2}\right)\left(\frac{\vct{a}_{1}^2}{r_{12}^4}-\frac{5(\vct{a}_{1}\!\cdot\!\vct{n}_{12})^2}{r_{12}^4}\right)\,,
\end{equation}

\begin{equation}
 H_{S_{2}^3p_1}=H_{S_{1}^3p_2}\,\left(1\leftrightarrow2\right),
\end{equation}

\begin{equation}
H_{S_{2}S_{1}^2p_2}=\;-G\frac{3m_{2}}{4m_{1}}\left(\frac{3\vct{a}_{1}^{2}\,\vct{p}_{2}\!\cdot\!\left(\vct{S}_{2}\times\vct{n}_{12}\right)}{r_{12}^4}+\frac{6(\vct{a}_{1}\!\cdot\!\vct{n}_{12})\,\vct{p}_{2}\!\cdot\!\left(\vct{S}_{2}\times\vct{a}_{1}\right)}{r_{12}^4}-\frac{15(\vct{a}_{1}\!\cdot\!\vct{n}_{12})^2\,\vct{p}_{2}\!\cdot\!\left(\vct{S}_{2}\times\vct{n}_{12}\right)}{r_{12}^4}\right)\,,
\end{equation}

\begin{equation}
 H_{S_{1}S_{2}^2p_1}=H_{S_{2}S_{1}^2p_2}\,\left(1\leftrightarrow2\right).
\end{equation}
The Hamiltonian $H_{S^4}$ splits into the parts 

\begin{equation}
 H_{S^4}=H_{S_{1}^2S_{2}^2} + H_{S_{1}S_{2}^3} +  H_{S_{1}^3S_{2}} + H_{S_{1}^4}+H_{S_{2}^4}.
\end{equation}
The Hamiltonians $H_{S_{1}^2S_{2}^2} + H_{S_{1}S_{2}^3} +
H_{S_{1}^3S_{2}}$ are given in \cite{HS} and the Hamiltonians
$H_{S_{1}^4}+H_{S_{2}^4}$ will be shown to be zero in Sec. VI below.
The remaining Hamiltonians will be chosen in the form, with to be
determined $\mu$-coefficients, 

\begin{align}
 \begin{aligned}
  H_{S_{1}^3p_1}=\frac{G}{r_{12}^4}\Bigg[\vct{S}_{1}\cdot\left(\vct{n}_{12}\times\vct{p}_{1}\right)\left(\mu_{1}\vct{S}_{1}^2+\mu_{2}\left(\vct{S}_{1}\cdot\vct{n}_{12}\right)^2\right)\Bigg],
 \end{aligned}
\end{align}

\begin{equation}
 H_{S_{2}^3p_2}=H_{S_{1}^3p_1}\,\left(1\leftrightarrow2\right),
\end{equation}

\begin{align}
 \begin{aligned}
 H_{S_{1}^2S_{2}p_1}=&\frac{G}{r_{12}^4}\bigg[\mu_{3}\vct{S}_{1}^2 \vct{S}_{2}\cdot\left(\vct{n}_{12}\times\vct{p}_{1}\right)+\mu_{4}\left(\vct{S}_{1}\cdot\vct{n}_{12}\right)\vct{S}_{2}\cdot\left(\vct{S}_{1}\times\vct{p}_{1}\right)+\mu_{5}\left(\vct{S}_{1}\cdot\vct{n}_{12}\right)^2\vct{S}_{2}\cdot\left(\vct{n}_{12}\times\vct{p}_{1}\right)\\
&\quad\quad+\mu_{6}\left(\vct{S}_{1}\cdot\vct{S}_{2}\right)\vct{S}_{1}\cdot\left(\vct{n}_{12}\times\vct{p}_{1}\right)+\mu_{7}\left(\vct{S}_{1}\cdot\vct{n}_{12}\right)\left(\vct{S}_{2}\cdot\vct{n}_{12}\right)\vct{S}_{1}\cdot\left(\vct{n}_{12}\times\vct{p}_{1}\right)\\
&\quad\quad+\vct{n}_{12}\cdot\left(\vct{S}_{1}\times\vct{S}_{2}\right)\big(\mu_{8}\vct{S}_{1}\cdot\vct{p}_{1}+\mu_{9}\left(\vct{S}_{1}\cdot\vct{n}_{12}\right)\left(\vct{p}_{1}\cdot\vct{n}_{12}\right)\big)\bigg],
 \end{aligned}
\end{align}

\begin{equation}
 H_{S_{2}^2S_{1}p_2}=H_{S_{1}^2S_{2}p_1}\,\left(1\leftrightarrow2\right).
\end{equation}
Notice that not all terms are independent of each other. They are
connected via the following identity for the mixed product of three vectors in three dimensions,

\begin{eqnarray}
 \left(U_{1}\,,U_{2}\,,U_{3}\right)\vct{U}&=&\left(UU_{1}\right)\vct{U}_{2}\times\vct{U}_{3}+\left(UU_{2}\right)\vct{U}_{3}\times\vct{U}_{1}+\left(UU_{3}\right)\vct{U}_{1}\times\vct{U}_{2}\\
                                          &=&\left(U\,,U_{2}\,,U_{3}\right)\vct{U}_{1}+\left(U_{1}\,,U\,,U_{3}\right)\vct{U}_{2}+\left(U_{1}\,,U_{2}\,,U\right)\vct{U}_{3}
\end{eqnarray}
with definitions $(UU_{1})=\vct{U}\cdot\vct{U}_{1}$ and
$(U_{1}\,,U_{2}\,,U_{3})=\vct{U}_{1}\cdot(\vct{U}_{2}\times\vct{U}_{3})$.
This will help us later to understand that some of the $\mu$ coefficients must be zero, because some terms can be shifted into others.
The ansatz for the $S_{1}^2p^2$ Hamiltonian reads, using $\alpha$,
$\beta$, and $\gamma$ coefficients, 

\begin{align}\label{p2S2Ham}
\begin{aligned}
 H_{S_{1}^2p^2}=&\frac{G}{r_{12}^3}\Bigg[\alpha_{1}\left(\vct{p}_{1}\cdot\vct{S}_{1}\right)^2+\alpha_{2}\vct{p}_{1}^{2}\vct{S}_{1}^{2}+\alpha_{3}\left(\vct{p}_{1}\cdot\vct{n}_{12}\right)^{2}\vct{S}_{1}^{2}+\alpha_{4}\vct{p}_{1}^{2}\left(\vct{S}_{1}\cdot\vct{n}_{12}\right)^2\\
&\qquad +\alpha_{5}\left(\vct{p}_{1}\cdot\vct{n}_{12}\right)^{2}\left(\vct{S}_{1}\cdot\vct{n}_{12}\right)^{2}+\alpha_{6}\left(\vct{p}_{1}\cdot\vct{n}_{12}\right)\left(\vct{S}_{1}\cdot\vct{n}_{12}\right)\left(\vct{p}_{1}\cdot\vct{S}_{1}\right)\\\\
&\qquad +\beta_{1}\left(\vct{p}_{2}\cdot\vct{S}_{1}\right)^2+\beta_{2}\vct{p}_{2}^{2}\vct{S}_{1}^{2}+\beta_{3}\left(\vct{p}_{2}\cdot\vct{n}_{12}\right)^{2}\vct{S}_{1}^{2}+\beta_{4}\vct{p}_{2}^{2}\left(\vct{S}_{1}\cdot\vct{n}_{12}\right)^2\\
&\qquad +\beta_{5}\left(\vct{p}_{2}\cdot\vct{n}_{12}\right)^{2}\left(\vct{S}_{1}\cdot\vct{n}_{12}\right)^{2}+\beta_{6}\left(\vct{p}_{2}\cdot\vct{n}_{12}\right)\left(\vct{S}_{1}\cdot\vct{n}_{12}\right)\left(\vct{p}_{2}\cdot\vct{S}_{1}\right)\\\\
&\qquad +\gamma_{1}\left(\vct{p}_{1}\cdot\vct{p}_{2}\right)\vct{S}_{1}^2+\gamma_{2}\left(\vct{p}_{1}\cdot\vct{p}_{2}\right)\left(\vct{S}_{1}\cdot\vct{n}_{12}\right)^2+\gamma_{3}\left(\vct{p}_{1}\cdot\vct{S}_{1}\right)\left(\vct{p}_{2}\cdot\vct{S}_{1}\right)\\
&\qquad
+\gamma_{4}\left(\vct{p}_{1}\cdot\vct{n}_{12}\right)\left(\vct{p}_{2}\cdot\vct{S}_{1}\right)\left(\vct{S}_{1}\cdot\vct{n}_{12}\right)+\gamma_{5}\left(\vct{p}_{2}\cdot\vct{n}_{12}\right)\left(\vct{p}_{1}\cdot\vct{S}_{1}\right)\left(\vct{S}_{1}\cdot\vct{n}_{12}\right)\\
&\qquad
+\gamma_{6}\left(\vct{p}_{1}\cdot\vct{n}_{12}\right)\left(\vct{p}_{2}\cdot\vct{n}_{12}\right)\vct{S}_{1}^2+\gamma_{7}\left(\vct{p}_{1}\cdot\vct{n}_{12}\right)\left(\vct{p}_{2}\cdot\vct{n}_{12}\right)\left(\vct{S}_{1}\cdot\vct{n}_{12}\right)^2\Bigg],
\end{aligned}
\end{align}

\begin{equation}
 H_{S_{2}^2p^2}=H_{S_{1}^2p^2} \,\left(1\leftrightarrow2\right).
\end{equation}

The center-of-mass vector $\vct{G}$ which enters in Eq. (\ref{crucial}) is given by

\begin{equation}
 \vct{G}=\vct{G}_{N}+\vct{G}_{1PN}+\vct{G}_{2PN}+\vct{G}_{SO}^{1PN}+\vct{G}_{SO}^{2PN}+\vct{G}_{S_{1}S_{2}}+\vct{G}_{S_{1}^2}
+ \vct{G}_{S_{2}^2}.
 \end{equation}
The point particle expressions $\vct{G}_{N}$, $\vct{G}_{1PN}$ and
$\vct{G}_{2PN}$ are given in \cite{DJS00}. The parts
$\vct{G}_{SO}^{1PN}$ and $\vct{G}_{SO}^{2PN}$ are presented in
\cite{DJS} and $\vct{G}_{S_{1}S_{2}}$ was calculated in \cite{SSH}. What
is left is to give the expression for $\vct{G}_{S_{1}^2}$ and
$\vct{G}_{S_{2}^2}$. Here in this section we will also make a general gauge invariant ansatz for them to show how they are involved in the algebra. It reads with coefficients $\nu_{1}$, $\nu_{2}$, $\nu_{3}$, $\nu_{4}$ and $\nu_{5}$

\begin{equation}
 \vct{G}_{S_{1}^2}=G\frac{m_{2}}{m_{1}}\left[\nu_{1}\frac{\left(\vct{S}_{1}\cdot\vct{n}_{12}\right)\vct{S}_1}{r_{12}^2}+\frac{\left(\vct{S}_{1}\cdot\vct{n}_{12}\right)^2}{r_{12}^3}\left(\nu_{2}\vct{x}_{1}+\nu_{3}\vct{x}_{2}\right)+\frac{\vct{S}_{1}^2}{r_{12}^3}\left(\nu_{4}\vct{x}_{1}+\nu_{5}\vct{x}_{2}\right)\right]
\end{equation}
plus the expression with $\left(1\leftrightarrow2\right)$.

Later we will calculate $\vct{G}_{S_{1}^2}$ directly by an appropriate covariant source for the Hamilton Constraint, which will fixate all the $\nu$-coefficients.



\section{Fulfillment of the Poincar\'e Algebra}

We concentrate on Eq. (\ref{crucial}) which gives rise to the
following equations for the coefficients:

\begin{eqnarray}
 \mu_{1}&=&\frac{m_{2}}{4m_{1}^3}\label{m1},\\
 \mu_{2}&=&-\frac{5}{4}\frac{m_{2}}{m_{1}^3},\\
 \mu_{3}&=&\frac{3}{m_{1}^2}-\mu_{8},\\
 \mu_{4}&=&\frac{15}{2m_{1}^2}+\mu_{9}+\mu_{8},\\
 \mu_{5}&=&-\frac{15}{m_{1}^2}-\mu_{9},\\
 \mu_{6}&=&-\frac{3}{2m_{1}^2}+\mu_{8},\\
 \mu_{7}&=&\frac{15}{2m_{1}^2}+\mu_{9},
\end{eqnarray}

\begin{eqnarray}
 \alpha_{1}&=&-\frac{3}{4}\frac{m_{2}}{m_{1}^3}-\frac{m_{2}}{2m_{1}}\gamma_{3}-\frac{m_{2}}{2m_{1}^3}\nu_{1}\label{a1},\\
 \alpha_{2}&=&\frac{m_{2}}{m_{1}^3}-\frac{m_{2}}{2m_{1}}\gamma_{1}+\frac{m_{2}}{2m_{1}^3}\nu_{5},\\
 \alpha_{3}&=&-\frac{9}{8}\frac{m_{2}}{m_{1}^3}-\frac{m_{2}}{2m_{1}}\gamma_{6}-\frac{3}{2}\frac{m_{2}}{m_{1}^3}\nu_{5},\\
 \alpha_{4}&=&-\frac{9}{8}\frac{m_{2}}{m_{1}^3}-\frac{m_{2}}{2m_{1}}\gamma_{2}-\frac{m_{2}}{2m_{1}^3}\nu_{2},\\
 \alpha_{5}&=&-\frac{15}{4}\frac{m_{2}}{m_{1}^3}-\frac{m_{2}}{2m_{1}}\gamma_{7}+\frac{5}{2}\frac{m_{2}}{m_{1}^3}\nu_{2},\\
 \alpha_{6}&=&\frac{15}{4}\frac{m_{2}}{m_{1}^3}-\frac{m_{2}}{m_{1}}\gamma_{4}+\frac{3m_{2}}{m_{1}^3}\nu_{1},\\
 \beta_{1}&=&\frac{1}{m_{1}m_{2}}-\frac{m_{1}}{2m_{2}}\gamma_{3}+\frac{\nu_{1}}{2m_{1}m_{2}},\\
 \beta_{2}&=&-\frac{1}{m_{1}m_{2}}-\frac{m_{1}}{2m_{2}}\gamma_{1}-\frac{\nu_{5}}{2m_{1}m_{2}},\\
 \beta_{3}&=&\frac{9}{4m_{1}m_{2}}-\frac{m_{1}}{2m_{2}}\gamma_{6}+\frac{3}{2m_{1}m_{2}}\nu_{5},\\
 \beta_{4}&=&\frac{3}{4m_{1}m_{2}}-\frac{m_{1}}{2m_{2}}\gamma_{2}+\frac{\nu_{2}}{2m_{1}m_{2}},\\
 \beta_{5}&=&-\frac{m_{1}}{2m_{2}}\gamma_{7}-\frac{5}{2m_{1}m_{2}}\nu_{2},\\
 \beta_{6}&=&-\frac{3}{m_{1}m_{2}}-\frac{m_{1}}{m_{2}}\gamma_{4}+\frac{2\nu_{2}}{m_{1}m_{2}},\\
 \gamma_{5}&=&\gamma_{4}-\frac{3\nu_{1}}{m_{1}^2}-\frac{2\nu_{2}}{m_{1}^2},\\
 \nu_{3}&=&\frac{3}{2}-\nu_{2},\\
 \nu_{4}&=&-\frac{1}{2}-\nu_{5}
 \label{g4}.
\end{eqnarray}

The Poincar\'e algebra obviously fixes the Hamiltonian $H_{S^3_1p_1}$
(and thus also  $H_{S^3_2p_2}$). It also restricts five from seven
coefficients for  $H_{S^2_1S_2p_1}$ (and thus also for $H_{S^2_2S_1p_2}$),
and two out of five coefficients for the center-of-mass vector
$\vct{G}_{S_{1}^2}$.
At this stage of our investigations, the coefficients in $H_{S^2_1p^2}$
and ${\bf G}_{S^2_1}$  can still be changed by the application of a canonical
transformation generated by  

\begin{align}
\begin{aligned}
g_{S^2_1p}=&\frac{G}{r_{12}^2}\Bigg(\vct{S}_{1}^2\left(\sigma_{1}\vct{p}_{1}\cdot\vct{n}_{12}+\sigma_{2}\vct{p}_{2}\cdot\vct{n}_{12}\right)+\vct{S}_{1}\cdot\vct{n}_{12}\left(\sigma_{3}\vct{S}_{1}\cdot\vct{p}_{1}+\sigma_{4}\vct{S}_{1}\cdot\vct{p}_{2}\right)\\
            &\qquad +
            \left(\vct{S}_{1}\cdot\vct{n}_{12}\right)^2\left(\sigma_{5}\vct{p}_{1}\cdot\vct{n}_{12}+\sigma_{6}\vct{p}_{2}\cdot\vct{n}_{12}\right)\Bigg) \label{generator}
\end{aligned}
\end{align}
with coefficients $\sigma_k$, $k=1,2,3,4,5,6$. Later on we shall see
that two coefficients of the center-of-mass vector $\vct{G}_{S_{1}^2}$,
and thus many other coefficients which are connected with them,  
can uniquely only be fixed with the aid of the explicit expression for the energy
density of a static source. Coming back to the Poincar\'e algebra, all its other
equations are trivially fulfilled and impose no further restrictions on
our coefficients. 


\section{The ADM generalized isotropic coordinates representation}

To get a unique representation of the Hamiltonians, we now have to fix
the coordinates. We will use the ADM formalism and generalized isotropic
coordinates. This makes it very easy to calculate interaction
Hamiltonians if the source expressions in the
constraint equations are known. Before imposing constraint equations and
coordinate conditions, the Hamiltonian in the ADM framework reads
\cite{ADM} \cite{RT}, 

\begin{equation}
 H=\int\dd^3 x(N\mathcal{H}-N^{i}\mathcal{H}_{i})+E\left[\gamma_{ij}\right],
\end{equation}
where respectively $N$ and $N^{i}$ denote lapse and shift function,
which are merely Lagrange multipliers.
The super-Hamiltonian $\mathcal{H}$ and the supermomentum
$\mathcal{H}_{i}$ densities decompose
into gravitational field and matter parts as follows:

\begin{equation}
 \mathcal{H}=\mathcal{H}^{\text{field}}+\mathcal{H}^{\text{matter}}\,,\quad\mathcal{H}_{i}=\mathcal{H}_{i}^{\text{field}}+\mathcal{H}_{i}^{\text{matter}}\,,
\end{equation}
where the field parts are given by

\begin{equation}
 \mathcal{H}^{\text{field}}=-\frac{1}{\sqrt{\gamma}}\left[\gamma
 \text{R} +\frac{1}{2}\left(\gamma_{ij}\pi^{ij}\right)^2-\gamma_{ij}\gamma_{kl}\pi^{ik}\pi^{jl}\right]\,,
 \quad\mathcal{H}_{i}^{\text{field}}=2\gamma_{ij}\pi^{jk}_{;k}.
\end{equation}

Here $\gamma$ is the
determinant of the 3-metric $\gamma_{ij} = g_{ij}$ of the spacelike
hypersurfaces $t = \text{const.}$, whereas the determinant of the
4-dimensional metric $g_{\mu\nu}$ will be denoted $g$.
The canonical conjugate to $\gamma_{ij}$ is $ \pi^{ij}$.
$\text{R}$ is the Ricci scalar of the spacelike hypersurfaces and $;$
denotes the 3-dimensional covariant derivative. The expressions for lapse and shift
functions in terms of the metric are $N=(-g^{00})^{-1/2}$ and
$N^i=\gamma^{ij}g_{0j}$, where $\gamma^{ij}$ is the inverse 3-metric.
As in \cite{SSH}, we will assume that the relation between field momentum $\pi^{ij}$ and
extrinsic curvature $K_{ij}$ is the same as in the vacuum case:
\begin{equation}\label{def_pi}
	\pi^{ij} = - \sqrt{\gamma} (\gamma^{ik}\gamma^{jl} - \gamma^{ij}\gamma^{kl})K_{kl}.
\end{equation}
The energy $E$ is defined by (herein, the comma ``,'' denotes partial 3-dimensional derivative)
\begin{equation}\label{ADM_energy}
E = \oint d^2 s_i (\gamma_{ij,j} -\gamma_{jj,i}) \,.
\end{equation}
$E$ turns into the ADM Hamiltonian $H_{\rm ADM}$ after imposing the
constraint equations $ \mathcal{H} =  \mathcal{H}_i  = 0$ and appropriate coordinate conditions.
Comparing the constraint equations with the Einstein field equations, projected onto
the spacelike hypersurfaces, results in
\begin{equation}\label{Hmatter}
{\cal{H}}^{\rm matter}= \sqrt{\gamma}T^{\mu\nu}n_{\mu}n_{\nu} =
N\sqrt{-g}T^{00} \, , \qquad
{\cal{H}}^{\rm matter}_i= - \sqrt{\gamma}T^{\nu}_i n_{\nu} =
\sqrt{-g}T^{0}_i \, ,
\end{equation}
where $\sqrt{-g}T^{\mu\nu}$ is the stress-energy tensor density of the
matter system. The timelike unit 4-vector $n_{\mu} = (-N,0,0,0)$ points
orthogonal to the spacelike hypersurfaces.

The generalized isotropic coordinates \cite{ADM}, also called the ADMTT gauge,
are the most often used and best adapted coordinate conditions for
explicit calculations and they are defined by
\begin{equation}\label{cc}
\gamma_{ij} = \psi^4 \delta_{ij} + h^{\text{TT}}_{ij}\quad\text{with}\quad\psi=1+\frac{1}{8}\phi\,, \quad \text{or} \quad
3\gamma_{ij,j} - \gamma_{jj,i} = 0 \,, \quad \text{and} \quad
\pi^{ii} = 0,  \quad \pi^{ij} = \tilde{\pi}^{ij} + \pi_{\text{TT}}^{ij}\,.
\end{equation}
$h^{\text{TT}}_{ij}$ has the properties  $ h^{\text{TT}}_{ij,j}  = h^{\text{TT}}_{ii} = 0$
(transverse and traceless) and the same holds for $\pi_{\text{TT}}^{ij}$
which is  the canonically conjugate to
$h^{\text{TT}}_{ij}$. $\tilde{\pi}^{ij}$ denotes the longitudinal part of  $\pi^{ij}$.

The ADM Hamiltonian and the center-of-mass vector read
\begin{equation}\label{conserved2}
	E = H_{\rm ADM} = -  \int{ d^3 x \, \Delta \phi } \,, \qquad
	G_i = -  \int{ d^3 x \, x^i \Delta \phi }\,,
\end{equation}
where $\phi$ is expressed in terms
of the canonical matter variables and the canonical field variables
$h^{\text{TT}}_{ij}$ and $\pi_{\text{TT}}^{ij}$ of the field-reduced phase
space. Herewith the Poincar\'e algebra can be verified. The most elegant
way for the obtention of an autonomous conservative Hamiltonian in the
matter variables is via the Routhian approach, see,
e.g. \cite{JS98}. However, to the order of our investigations 
$\pi_{\text{TT}}^{ij}$ does not play any role so we may stay on the
Hamiltonian level for the field variables. 

The constraint equations explicitly read

\begin{eqnarray}
\frac{1}{\sqrt{\gamma}}\left[\gamma \text{R}+\frac{1}{2}\left(\gamma_{ij}\pi^{ij}\right)^2-\gamma_{ij}\gamma_{kl}\pi^{ik}\pi^{jl}\right]&=&\mathcal{H}^{\text{matter}},\\
 -2\gamma_{ij}\pi^{jk}_{;k}&=&\mathcal{H}_{i}^{\text{matter}}.
\end{eqnarray}

To include momentum squared terms interacting with spin-squared terms we
have to make an ansatz for the sources generalizing the point particle
terms without spin. The most general ansatz for our purpose reads for
the source in the Hamilton constraint, cf. \cite{HS}, 

\begin{align}
\begin{aligned}
 \mathcal{H}^{\text{matter}}=&\sum_{b}\Bigg[-\frac{m_{b}}{2}Q_{b}^{ij}\partial_{i}\partial_{j}-\frac{1}{2}\vct{p}_{b}\cdot\left(\vct{a}_{b}\times\vct{\partial}\right)+\left(\gamma^{ij}p_{bi}p_{bj}+m_{b}^2\right)^{1/2}+\lambda_{1}\frac{\vct{p}_{b}^2}{2m_{b}}Q_{b}^{ij}\partial_{i}\partial_{j}\\
                             &\quad+\frac{\lambda_{2}}{m_{b}}(\vct{p}_{b}\cdot\vct{\partial})Q_{b}^{ij}p_{bi}\partial_{j}+\frac{\lambda_{3}}{m_{b}}\vct{a}_{b}^2(\vct{p}_{b}\cdot\vct{\partial})^2-\lambda_{8}\vct{p}_{b}\cdot\left(\vct{a}_{b}\times\vct{\partial}\right)Q_{b}^{ij}\partial_{i}\partial_{j}\Bigg]\delta_{b}
\end{aligned}
\end{align}
with the quadrupole tensor

\begin{equation}
 Q_{b}^{ij}=a_{b}^{i}a_{b}^{j}-\frac{1}{3}\vct{a}^2_{b}\delta_{ij}\,,\label{quadrupole}
\end{equation}
where by definition $\vct{a}_{b}=\vct{S}_{b}/m_b$ holds and $\delta_b
\equiv \delta(x^i - x^i_b)$ with $\int d^3x~ \delta_b = 1$ and $\vct{\partial}=
(\partial_i), \partial_i = \frac{\partial}{\partial x^i}$.
Notice that purely Laplacian source terms, $\Delta \delta_b$, apart from the ones appearing
via $Q^{ij}$, are left out because they would lead to a distributional
Hamiltonian (cf. paragraph on Breit's equation in \cite{LL}) which is of no interest here.

The ansatz for the source of the momentum constraint reads, also not taking into
account terms of the form $\Delta \delta_b$ apart from the ones
appearing with $Q^{ij}$ (taking into account those terms would lead to no
effects on our Hamiltonians when controlled by Poincar\'e algebra),
cf. \cite{HS},

\begin{align}
\begin{aligned}
 \mathcal{H}^{\text{matter}}_{i}=&-2\sum_{b}\Bigg[Q_{b}^{kl}\bigg(\lambda_{5}p_{bk}\partial_{l}\partial_{i}+\lambda_{6}p_{bi}\partial_{k}\partial_{l}
+\lambda_{7}\left(\vct{p}_{b}\cdot\vct{\partial}\right)\delta_{li}\partial_{k}\bigg)+\lambda_{4}\vct{a}_{b}^2(\vct{p}_{b}\cdot\vct{\partial})\partial_{i}\\
&+\frac{m_{b}}{4}\left(\vct{a}_{b}\times\vct{\partial}\right)_{i}\left(1-\frac{1}{6}Q_{b}^{kl}\partial_{k}\partial_{l}\right)-\frac{1}{2}p_{bi}\Bigg]\delta_{b}.
\end{aligned}
\end{align}
This ansatz is not 3-dimensional general covariant, but it is of sufficient general form that
will lead to all the searched for contributions of the $H_{S^3p}$ and 
$H_{S^2p^2}$ Hamiltonians. It also correctly holds
$\int  d^3x~ \mathcal{H}^{\text{matter}}_{i} = P_i$ and 
$\int  d^3x~ \epsilon_{ijk} x^j\mathcal{H}^{\text{matter}}_k = J_i$.
To the order needed for a 2PN-Hamiltonian for
self-spin interaction, the Hamilton constraint expands as
\begin{align}
	-  \Delta \phi_{(2)} &= \mathcal{H}^{\rm matter}_{(2)}\,, \qquad
	-  \Delta \phi_{(4)} = \mathcal{H}^{\rm matter}_{(4)}
		- \frac{1}{8} \mathcal{H}^{\rm matter}_{(2)} \phi_{(2)}\,, \\
	-  \Delta \phi_{(6)} &= \mathcal{H}^{\rm matter}_{(6)}
		- \frac{1}{8} ( \mathcal{H}^{\rm matter}_{(4)} \phi_{(2)}
		+ \mathcal{H}^{\rm matter}_{(2)} \phi_{(4)} )
		+ \frac{1}{64} \mathcal{H}^{\rm matter}_{(2)} \phi_{(2)}^2
		+  \left[ ( \tilde{\pi}^{i j}_{(3)} )^2
		- \frac{1}{2} \phi_{(2) , i j} h^{\text{TT}}_{(4) i j} \right]
		\,, \label{phi6_const} \\
\begin{split}
	-  \Delta \phi_{(8)} &= \mathcal{H}^{\rm matter}_{(8)}
		- \frac{1}{8} ( \mathcal{H}^{\rm matter}_{(6)} \phi_{(2)}
			+ \mathcal{H}^{\rm matter}_{(4)} \phi_{(4)}
			+ \mathcal{H}^{\rm matter}_{(2)} \phi_{(6)} )
		+ \frac{1}{64} ( \mathcal{H}^{\rm matter}_{(4)} \phi_{(2)}^2
		+ 2 \mathcal{H}^{\rm matter}_{(2)} \phi_{(2)} \phi_{(4)} ) \\
	&\quad	- \frac{1}{512} \mathcal{H}^{\rm matter}_{(2)} \phi_{(2)}^3
		+ \left[ \frac{1}{8} \phi_{(2)} (\tilde{\pi}^{i j}_{(3)})^2
		+ 2 \tilde{\pi}^{i j}_{(3)} \tilde{\pi}^{i j}_{(5)}
		- \frac{1}{16} \phi_{(2) , i} \phi_{(2) , j} h^{\text{TT}}_{(4) i j}
		+ \frac{1}{4} (h^{\text{TT}}_{(4) i j , k})^2 \right]
		+ (\text{td}) \,,
\end{split}
\end{align}
(``td'' means total derivative) and the momentum constraint can be written as
\begin{align}
	\tilde{\pi}^{i j}_{(3) , j} &=
		- \frac{1}{2} \mathcal{H}^{\rm matter}_{(3) i}\,, \\
	\tilde{\pi}^{i j}_{(5) , j} &=
		- \frac{1}{2} \mathcal{H}^{\rm matter}_{(5) i}
		- \frac{1}{2} ( \phi_{(2)} \tilde{\pi}^{i j}_{(3)} )_{, j} \,.\label{pi5_const}
\end{align}
The integral over $-\Delta\phi_{(8)}$ is the one leading to the new
Hamiltonians to linear order in G. The $h_{ij}^{TT}$ term will drop out
from the calculation of the Hamiltonians in question. 
The expressions needed for this integral read 

\begin{eqnarray}
 \mathcal{H}^{\text{matter}}_{(2)}&=&\sum_{b}m_{b}\delta_{b},\\
 \mathcal{H}^{\text{matter}}_{(4)}&=&\sum_{b}\Bigg[-\frac{m_{b}}{2}Q_{b}^{ij}\partial_{i}\partial_{j}+\frac{\vct{p}_{b}^2}{2m_{b}}-\frac{1}{2}\vct{p}_{b}\cdot\left(\vct{a}_{b}\times\vct{\partial}\right)\Bigg]\delta_{b},\\
 \mathcal{H}^{\text{matter}}_{(6)}&=&\sum_{b}\Bigg(-\frac{1}{4}\phi_{(2)}\frac{\vct{p}_{b}^2}{m_{b}}-\lambda_{8}\vct{p}_{b}\cdot\left(\vct{a}_{b}\times\vct{\partial}\right)Q_{b}^{ij}\partial_{i}\partial_{j}+\lambda_{1}\frac{\vct{p}_{b}^2}{2m_{b}}Q_{b}^{ij}\partial_{i}\partial_{j}\nonumber\\
& &\quad+\frac{\lambda_{2}}{m_{b}}(\vct{p}_{b}\cdot\vct{\partial})Q_{b}^{ij}p_{bi}\partial_{j}+\frac{\lambda_{3}}{m_{b}}\vct{a}_{b}^2(\vct{p}_{b}\cdot\vct{\partial})^2\Bigg)\delta_{b},\\
 \mathcal{H}^{\text{matter}}_{(8)}&=&\sum_{b} \left(-\frac{1}{4}\phi_{(4)}\frac{\vct{p}^2_{b}}{m_{b}}+\frac{5}{64}\phi_{(2)}^2\frac{\vct{p}^2_{b}}{m_{b}}\right),\\
\mathcal{H}^{\text{matter}}_{(3)i}&=&\sum_{b}\left(p_{bi}-\frac{1}{2}m_{b}\left(\vct{a}_{b}\times\vct{\partial}\right)_{i}\right)\delta_{b},\\
\mathcal{H}^{\text{matter}}_{(5)i}&=&-2\sum_{b}\Bigg[Q_{b}^{kl}\bigg(\lambda_{5}p_{bk}\partial_{l}\partial_{i}+\lambda_{6}p_{bi}\partial_{k}\partial_{l}+\lambda_{7}\left(\vct{p}_{b}\cdot\vct{\partial}\right)\delta_{li}\partial_{k}\bigg)\\
&&\quad+\lambda_{4}\vct{a}_{b}^2(\vct{p}_{b}\cdot\vct{\partial})\partial_{i}
+\frac{m_{b}}{24}\left(\vct{a}_{b}\times\vct{\partial}\right)_{i}Q_{b}^{kl}\partial_{k}\partial_{l}\Bigg]\delta_{b},
\end{eqnarray}

and

\begin{eqnarray}
 \phi_{(2)}&=&4G\sum_{b}\frac{m_{b}}{r_{b}},\\
 \phi_{(4)}&=&4G\sum_{b}\Bigg[-\frac{m_{b}}{2}Q_{b}^{ij}\partial_{bi}\partial_{bj}+\frac{\vct{p}_{b}^2}{2m_{b}}+\frac{1}{2}\vct{p}_{b}\cdot\left(\vct{a}_{b}\times\vct{\partial}_{b}\right)\Bigg]\frac{1}{r_{b}}-2G^2\sum_{a \ne b}m_{a}m_{b}\frac{1}{r_{ab}r_{b}},\\
 \phi_{(6)}^{\text{I}}&=&-\Delta^{-1}\mathcal{H}^{\rm matter}_{(6)}\nonumber\\
                      &=&4G\sum_{b}\Bigg(\lambda_{8}\vct{p}_{b}\cdot\left(\vct{a}_{b}\times\vct{\partial}_{b}\right)Q_{b}^{ij}\partial_{bi}\partial_{bj}+\lambda_{1}\frac{\vct{p}_{b}^2}{2m_{b}}Q_{b}^{ij}\partial_{bi}\partial_{bj}+\frac{\lambda_{2}}{m_{b}}(\vct{p}_{b}\cdot\vct{\partial}_{b})Q_{b}^{ij}p_{bi}\partial_{bj}+\frac{\lambda_{3}}{m_{b}}\vct{a}_{b}^2(\vct{p}_{b}\cdot\vct{\partial}_{b})^2\Bigg)\frac{1}{r_{b}}\nonumber\\
           & &-\sum_{a \ne b}4G^2\frac{m_{a}}{m_{b}}\vct{p}_{b}^2\frac{1}{r_{ab}r_{b}},\\
 \phi_{(6)}^{\text{II}}&=&-\Delta^{-1}\left[-\frac{1}{8}\left(\mathcal{H}_{(4)}^{\text{matter}}\phi_{(2)}+\mathcal{H}_{(2)}^{\text{matter}}\phi_{(4)}\right)\right],\\
 \phi_{(6)}^{\text{III}}&=&-\Delta^{-1}\left(\frac{1}{64}\mathcal{H}_{(2)}^{\text{matter}}\phi_{(2)}^2\right),\\
 \phi_{(6)}^{\text{IV}}&=&-\Delta^{-1}\left(\tilde{\pi}^{ij}_{(3)}\right)^2,
\end{eqnarray}

\begin{eqnarray}
 \tilde{\pi}^{ij}_{(3)}&=&\Theta^{ij}_{k}\left(- \frac{1}{2} \mathcal{H}^{\rm matter}_{(3) k}\right),\\
 \tilde{\pi}^{ij}_{(5)}&=&\Theta^{ij}_{k}\left(- \frac{1}{2} \mathcal{H}^{\rm matter}_{(5) k}
		- \frac{1}{2} ( \phi_{(2)} \tilde{\pi}^{k l}_{(3)} )_{, l}\right)\qquad\text{with}\\
 \Theta^{ij}_{k}& \equiv &\left(-\frac{1}{2}\delta_{ij}\partial_{k}+\delta_{ik}\partial_{j}+\delta_{jk}\partial_{i}-\frac{1}{2}\partial_{i}\partial_{j}\partial_{k}\Delta^{-1}\right)\Delta^{-1}\,.
\end{eqnarray}
We do not need to know $\tilde{\pi}^{ij}_{(3)}$ and
$\tilde{\pi}^{ij}_{(5)}$ in full detail as we can apply partial
integrations within our used analytical regularization procedure. 
The integrals are calculated with the aid of the same methods and
regularization formulas as outlined in, e.g. \cite{SSH}.
The results are (the symbol $\simeq$ indicates that only the relevant contributions are given)

\begin{eqnarray}
 H_{(8)1}&=&\int\mathcal{H}^{\text{matter}}_{(8)}\dd^3 x\simeq G\sum_{a
 \ne b}\frac{1}{2}\frac{m_{a}}{m_{b}}\vct{p}_{b}^2 Q_{a}^{ij}\partial_{ai}\partial_{aj}\frac{1}{r_{ab}},\\
 H_{(8)2}&=&\int-\frac{1}{8}\mathcal{H}_{(6)}^{\text{matter}}\phi_{(2)}\dd^3 x\nonumber\\
         &\simeq&-\frac{G}{2}\sum_{a \ne b}m_{a}\bigg(\lambda_{8}\vct{p}_{b}\cdot\left(\vct{a}_{b}\times\vct{\partial}_{b}\right)Q_{b}^{ij}\partial_{bi}\partial_{bj}+\lambda_{1}\frac{\vct{p}_{b}^2}{2m_{b}}Q_{b}^{ij}\partial_{bi}\partial_{bj}+\frac{\lambda_{2}}{m_{b}}(\vct{p}_{b}\cdot\vct{\partial}_{b})Q_{b}^{ij}p_{bi}\partial_{bj}\nonumber\\
& &\qquad\qquad\quad+\frac{\lambda_{3}}{m_{b}}\vct{a}_{b}^2(\vct{p}_{b}\cdot\vct{\partial}_{b})^2\bigg)\frac{1}{r_{ab}},\\
 H_{(8)3}&=&\int-\frac{1}{8}\mathcal{H}_{(2)}^{\text{matter}}\phi_{(6)}\dd^3 x=\int-\frac{1}{8}\mathcal{H}_{(2)}^{\text{matter}}\left(\phi_{(6)}^{\text{I}}+\phi_{(6)}^{\text{II}}+\phi_{(6)}^{\text{III}}+\phi_{(6)}^{\text{IV}}\right)\dd^3 x,\\
 H_{(8)3}^{\text{I}}&=&H_{(8)2},\\
 H_{(8)3}^{\text{II}}&=&\mathcal{O}\left(G^2\right),\\
 H_{(8)3}^{\text{III}}&=&\mathcal{O}\left(G^2\right),\\
 H_{(8)3}^{\text{IV}}&=&\int\frac{1}{8}\Delta\phi_{(2)}\phi_{(6)}^{\text{IV}}\dd^3 x=\int-\frac{1}{8}\phi_{(2)}\left(\tilde{\pi}^{ij}_{(3)}\right)^2\dd^3 x,\\
 H_{(8)4}&=&\int-\frac{1}{8}\mathcal{H}^{\text{matter}}_{(4)}\phi_{(4)}\dd^3 x\\
         &\simeq&\frac{1}{4}G\sum_{a \ne b}\frac{m_{a}}{m_{b}}\vct{p}_{b}^2 Q_{a}^{ij}\partial_{ai}\partial_{aj}\frac{1}{r_{ab}}+\mathcal{O}\left(G^2\right),\\
 H_{(8)5}&\simeq&\int\frac{1}{64}\mathcal{H}^{\text{matter}}_{(4)}\phi_{(2)}^2\dd^3 x=\mathcal{O}\left(G^2\right),\\
 H_{(8)8}&=&\int\frac{1}{8}\phi_{(2)}\left(\tilde{\pi}^{ij}_{(3)}\right)^2\dd^3 x=-H_{(8)3}^{\text{IV}},
\end{eqnarray}

\begin{align}
 \begin{aligned}
  H_{(8)9}=&\int 2\tilde{\pi}^{i j}_{(3)} \tilde{\pi}^{i j}_{(5)}\dd^3 x\\
\simeq&\frac{G}{m_{1}^2 r_{12}^3}\left(\lambda_{4}-\frac{1}{3}\lambda_{5}-\frac{1}{3}\lambda_{7}\right)\Bigg[6\vct{S}_{1}^2\left(\vct{p}_{1}\cdot\vct{p}_{2}\right)-18\vct{S}_{1}^2\left(\vct{p}_{1}\cdot\vct{n}_{12}\right)\left(\vct{p}_{2}\cdot\vct{n}_{12}\right)\Bigg]\\
&+\frac{G\lambda_{5}}{m_{1}^2 r_{12}^3}\Bigg[6\left(\vct{S}_{1}\cdot\vct{p}_{1}\right)\left(\vct{S}_{1}\cdot\vct{p}_{2}\right)-18\left(\vct{S}_{1}\cdot\vct{n}_{12}\right)\left(\vct{p}_{2}\cdot\vct{n}_{12}\right)\left(\vct{S}_{1}\cdot\vct{p}_{1}\right)\Bigg]\\
&+\frac{G\lambda_{6}}{m_{1}^2 r_{12}^3}\Bigg[-15\left(\vct{p}_{1}\cdot\vct{n}_{12}\right)\left(\vct{p}_{2}\cdot\vct{n}_{12}\right)\left(\vct{S}_{1}\cdot\vct{n}_{12}\right)^2-21\left(\vct{p}_{1}\cdot\vct{p}_{2}\right)\left(\vct{S}_{1}\cdot\vct{n}_{12}\right)^2\\
&\qquad+6\left(\vct{p}_{2}\cdot\vct{n}_{12}\right)\left(\vct{S}_{1}\cdot\vct{n}_{12}\right)\left(\vct{p}_{1}\cdot\vct{S}_{1}\right)+6\left(\vct{p}_{1}\cdot\vct{n}_{12}\right)\left(\vct{S}_{1}\cdot\vct{n}_{12}\right)\left(\vct{p}_{2}\cdot\vct{S}_{1}\right)\\
&\qquad-2\left(\vct{p}_{1}\cdot\vct{S}_{1}\right)\left(\vct{p}_{2}\cdot\vct{S}_{1}\right)+3\left(\vct{p}_{1}\cdot\vct{n}_{12}\right)\left(\vct{p}_{2}\cdot\vct{n}_{12}\right)\vct{S}_{1}^2+7\left(\vct{p}_{1}\cdot\vct{p}_{2}\right)\vct{S}_{1}^2\Bigg]\\
&+\frac{G\lambda_{7}}{m_{1}^2 r_{12}^3}\Bigg[-15\left(\vct{p}_{1}\cdot\vct{n}_{12}\right)\left(\vct{p}_{2}\cdot\vct{n}_{12}\right)\left(\vct{S}_{1}\cdot\vct{n}_{12}\right)^2+3\left(\vct{p}_{1}\cdot\vct{p}_{2}\right)\left(\vct{S}_{1}\cdot\vct{n}_{12}\right)^2\\
&\qquad+6\left(\vct{p}_{2}\cdot\vct{n}_{12}\right)\left(\vct{S}_{1}\cdot\vct{n}_{12}\right)\left(\vct{p}_{1}\cdot\vct{S}_{1}\right)-18\left(\vct{p}_{1}\cdot\vct{n}_{12}\right)\left(\vct{S}_{1}\cdot\vct{n}_{12}\right)\left(\vct{p}_{2}\cdot\vct{S}_{1}\right)\\
&\qquad+6\left(\vct{p}_{1}\cdot\vct{S}_{1}\right)\left(\vct{p}_{2}\cdot\vct{S}_{1}\right)+3\left(\vct{p}_{1}\cdot\vct{n}_{12}\right)\left(\vct{p}_{2}\cdot\vct{n}_{12}\right)\vct{S}_{1}^2-\left(\vct{p}_{1}\cdot\vct{p}_{2}\right)\vct{S}_{1}^2\Bigg]\\
& - \frac{G\lambda_{6}}{3m_{1}^2 r_{12}^3}\Bigg[-2\vct{S}_{1}^2\left(\vct{p}_{1}\cdot\vct{p}_{2}\right)+6\vct{S}_{1}^2\left(\vct{p}_{1}\cdot\vct{n}_{12}\right)\left(\vct{p}_{2}\cdot\vct{n}_{12}\right)\Bigg]\\
&+\frac{G\lambda_{6}}{m_{1}^2 r_{12}^4}\Bigg(-60\left(\vct{S}_{1}\cdot\vct{n}_{12}\right)^2\vct{n}_{12}\cdot\left(\vct{p}_{1}\times\vct{S}_{2}\right)+12\vct{S}_{1}^2\vct{n}_{12}\cdot\left(\vct{p}_{1}\times\vct{S}_{2}\right)+24\left(\vct{S}_{1}\cdot\vct{n}_{12}\right)\vct{p}_{1}\cdot\left(\vct{S}_{2}\times\vct{S}_{1}\right)\Bigg)\\
&+\frac{G\lambda_{7}}{m_{1}^2 r_{12}^4}\Bigg(-60\left(\vct{p}_{1}\cdot\vct{n}_{12}\right)\left(\vct{S}_{1}\cdot\vct{n}_{12}\right)\vct{n}_{12}\cdot\left(\vct{S}_{1}\times\vct{S}_{2}\right)+12\left(\vct{p}_{1}\cdot\vct{S}_{1}\right)\vct{n}_{12}\cdot\left(\vct{S}_{1}\times\vct{S}_{2}\right)\\
&\qquad\qquad\quad+12\left(\vct{S}_{1}\cdot\vct{n}_{12}\right)\vct{p}_{1}\cdot\left(\vct{S}_{1}\times\vct{S}_{2}\right)\Bigg)\\
&+(1\leftrightarrow2).
 \end{aligned}
\end{align}

We conclude

\begin{eqnarray}
 H_{(8)1}+H_{(8)4}&=&\frac{3}{4}\frac{G}{m_{1}m_{2}}\vct{p}_{2}^2\left(\frac{3(\vct{S}_{1}\cdot\vct{n}_{12})^2-\vct{S}_{1}^2}{r_{12}^3}\right)\,+\left(1\leftrightarrow2\right),\\
 H_{(8)2}+H_{(8)3}^{I}&=&-G\frac{m_{2}}{m_{1}^3}\Bigg[\lambda_{1}\frac{\vct{p}_{1}^2}{2}\frac{3(\vct{S}_{1}\cdot\vct{n}_{12})^2-\vct{S}_{1}^2}{r_{12}^3}+\left(\lambda_{3}-\frac{1}{3}\lambda_{2}\right)\vct{S}_{1}^2\frac{3(\vct{p}_{1}\cdot\vct{n}_{12})^2-\vct{p}_{1}^2}{r_{12}^3}\nonumber\\
& &\qquad\qquad+\lambda_{2}\frac{3(\vct{p}_{1}\cdot\vct{n}_{12})(\vct{S}_{1}\cdot\vct{n}_{12})(\vct{p}_{1}\cdot\vct{S}_{1})-(\vct{p}_{1}\cdot\vct{S}_{1})^2}{r_{12}^3}\nonumber\\
& &\qquad\qquad-\lambda_{8}\frac{(-3\vct{S}_{1}^2+15(\vct{S}_{1}\cdot\vct{n}_{12})^2)\vct{p}_{1}\cdot(\vct{S}_{1}\times\vct{n}_{12})}{r_{12}^4}\Bigg]\,+\left(1\leftrightarrow2\right),\\
 H_{(8)3}^{\text{IV}}+H_{(8)8}&=&0.
\end{eqnarray}

The coefficient equations following from our considerations read 

\begin{eqnarray}
 \alpha_{1}&=&\frac{m_{2}}{m_{1}^3}\lambda_{2},
 \quad\alpha_{2}=\frac{m_{2}}{2m_{1}^3}\left(\lambda_{1} + 2\lambda_{3}-\frac{2}{3}\lambda_{2}\right),
 \quad\alpha_{3}=-3\frac{m_{2}}{m_{1}^3}\left(\lambda_{3}-\frac{1}{3}\lambda_{2}\right),\\
 \alpha_{4}&=&-\frac{3}{2}\lambda_{1}\frac{m_{2}}{m_{1}^3},\quad\alpha_{5}=0,\quad\alpha_{6}=-3\lambda_{2}\frac{m_{2}}{m_{1}^2},\\
 \beta_{1}&=&\beta_{3}=\beta_{5}=\beta_{6}=0,\quad\beta_{2}=-\frac{3}{4m_{1}m_{2}},\quad\beta_{4}=\frac{9}{4m_{1}m_{2}},\\
 \gamma_{1}&=&\frac{1}{m_{1}^2}\left(6\lambda_{4}-2\lambda_{5}-3\lambda_{7}+\frac{23}{3}\lambda_{6}\right),\\
 \gamma_{2}&=&\frac{1}{m_{1}^2}\left(-21\lambda_{6}+3\lambda_{7}\right),\\
 \gamma_{3}&=&\frac{1}{m_{1}^2}\left(6\lambda_{5}-2\lambda_{6}+6\lambda_{7}\right),\\
 \gamma_{4}&=&\frac{1}{m_{1}^2}\left(6\lambda_{6}-18\lambda_{7}\right),\\
 \gamma_{5}&=&\frac{1}{m_{1}^2}\left(-18\lambda_{5}+6\lambda_{6}+6\lambda_{7}\right),\\
 \gamma_{6}&=&\frac{1}{m_{1}^2}\left(-18\lambda_{4}+6\lambda_{5}+9\lambda_{7}+\lambda_{6}\right),\\
 \gamma_{7}&=&\frac{1}{m_{1}^2}\left(-15\lambda_{6}-15\lambda_{7}\right),\\
 \mu_{1}&=&-3\frac{m_{2}}{m_{1}^3}\lambda_{8},\quad\mu_{2}=15\frac{m_{2}}{m_{1}^3}\lambda_{8},\quad\mu_{3}
 =\frac{12}{m_{1}^2}\lambda_{6},\\
 \mu_{4}&=&\frac{1}{m_{1}^2}\left(24\lambda_{6}-12\lambda_{7}\right),\quad\mu_{5}=\frac{-60}{m_{1}^2}\lambda_{6},\quad\mu_{8}=\frac{12}{m_{1}^2}\lambda_{7},\quad\mu_{9}=\frac{-60}{m_{1}^2}\lambda_{7},\\
 \mu_{6}&=&\mu_{7}=0.
\end{eqnarray}
Now we make use of the Eqs. (\ref{m1}) to (\ref{g4}) and end up with the solution for all the coefficients still having 2 degrees of freedom left parametrized by $\nu_{1}$ and $\nu_{5}$:

\begin{eqnarray}
 \alpha_{1}&=&\frac{m_{2}}{m_{1}^3}\left(-\frac{7}{4}-\nu_{1}\right),\quad\alpha_{2}=\frac{m_{2}}{m_{1}^3}\left(\frac{5}{4}+\nu_{5}\right),\quad\alpha_{3}=\frac{m_{2}}{m_{1}^3}\left(-\frac{27}{8}-3\nu_{5}\right),\quad\alpha_{4}=-\frac{3}{8}\frac{m_{2}}{m_{1}^3},\quad\alpha_{5}=0,\\
\alpha_{6}&=&\frac{m_{2}}{m_{1}^3}\left(\frac{21}{4}+3\nu_{1}\right),\\
 \beta_{1}&=&\beta_{3}=\beta_{5}=\beta_{6}=0,\quad\beta_{2}=-\frac{3}{4m_{1}m_{2}},\quad\beta_{4}=\frac{9}{4m_{1}m_{2}},\\
 \gamma_{1}&=&\frac{1}{m_{1}^2}\left(-1-\nu_{5}\right),\quad\gamma_{2}=-\frac{9}{4m_{1}^2},\quad\gamma_{3}=\frac{1}{m_{1}^2}\left(2+\nu_{1}\right),\quad\gamma_{4}=-\frac{3}{2m_{1}^2},\quad\gamma_{5}=\frac{1}{m_{1}^2}\left(-3-3\nu_{1}\right),\\
\gamma_{6}&=&\frac{1}{m_{1}^2}\left(\frac{9}{2}+3\nu_{5}\right),\quad\gamma_{7}=-\frac{15}{4m_{1}^2},\\
 \mu_{1}&=&\frac{m_{2}}{4m_{1}^3},\quad\mu_{2}=-\frac{5}{4}\frac{m_{2}}{m_{1}^3},\quad\mu_{3}=\frac{3}{2m_{1}^2},\quad\mu_{4}=\frac{3}{2m_{1}^2},\quad\mu_{5}=-\frac{15}{2m_{1}^2},\quad\mu_{8}=\frac{3}{2m_{1}^2},\quad\mu_{9}=-\frac{15}{2m_{1}^2},\\
 \mu_{6}&=&\mu_{7}=0\,,\\
 \nu_{2}&=&\frac{3}{4}, \quad\nu_{3}=\frac{3}{4}, \quad\nu_{4}=-\frac{1}{2}-\nu_{5}\label{nucoeff}
\end{eqnarray}

\begin{eqnarray}
 \lambda_{1}&=&\frac{1}{4}, \quad\lambda_{2}=-\frac{7}{4}-\nu_{1}, \quad\lambda_{3}=\frac{13}{24}-\frac{\nu_{1}}{3}+\nu_{5},
 \quad\lambda_{4}=-\frac{7}{72}+\frac{\nu_{1}}{18}-\frac{\nu_{5}}{6},
 \quad\lambda_{5}=\frac{1}{4}+\frac{\nu_{1}}{6},\quad\lambda_{6}=\frac{1}{8},\\
\lambda_{7}&=&\frac{1}{8},\quad\lambda_{8}=-\frac{1}{12}.
\end{eqnarray}

It is worthy to point out, that the free parameters $\nu_{1}$ and
$\nu_{5}$ are related to a canonical transformation which is given by the generator

\begin{equation}
 g_{S_1^2p_1} =\frac{Gm_{2}}{m_{1}^2r_{12}^2}\left(-\nu_{5}\vct{S}_{1}^2(\vct{p}_{1}\cdot\vct{n}_{12})+\nu_{1}(\vct{S}_{1}\cdot\vct{n}_{12})(\vct{p}_{1}\cdot\vct{S}_{1})\right)\,.
\end{equation}

Refering to Eq. (\ref{generator}) this means that we may choose
$\sigma_{1}=-\frac{m_2}{m_1^2}\nu_{5} $, $\sigma_{3}= \frac{m_2}{m_1^2}\nu_{1}$, and
$\sigma_{2}=\sigma_{4}= \sigma_{5}=\sigma_{6}=0$ to generate the terms related with these coefficients. So all Hamiltonians parametrized
by $\nu_{1}$ and $\nu_{5}$ are canonically equivalent and we are free to
give them any value. To fixate them we need to calculate
$\vct{G}_{S_{1}^2}$ explicitly. We use Eq. (\ref{conserved2}) and as a
matter source for the static case we adopt the result from \cite{SHS},
Eq. (4), where a static source expression for a black-hole binary has
been derived for the next-to-leading order spin-squared terms:

\begin{equation}
\begin{split}
	\mathcal{H}^{\rm matter}_{\text{S$_1^2$, static}} &=
		-\frac{1}{2m_{1}} \left( I^{ij}_{1} \delta_1 \right)_{; ij}
		+ \frac{c_3}{m_1} \vct{S}_1^2 \left( \gamma^{ij} \delta_1 \right)_{;ij}
                + \frac{1}{8 m_1} g_{mn} \gamma^{pj} \gamma^{ql} \gamma^{mi}_{~~,p} \gamma^{nk}_{~~,q} \hat{S}_{1 ij} \hat{S}_{1 kl} \delta_1 \nl
	        + \frac{1}{4m_1} \left( \gamma^{ij} \gamma^{mn} \gamma^{kl}_{~~,m} \hat{S}_{1 ln} \hat{S}_{1 jk} \delta_{1}\right)_{,i} \,
\end{split}
\end{equation}
with, in the present approximation, $I^{ij}_{1}=m_{1}^2\gamma^{jk}Q^{kj}_{1}$ and $\hat{S}_{1
ij}=\epsilon_{kij}S_{1k}$. Notice that contributions arising from the $c_{3}$ source term cancel each other, which is very nice, because $c_{3}$ could not be determined and does also not contribute to the $G^2S_{1}^2$ Hamiltonian.
The explicit calculation then yields $\nu_{1}=-2$, $\nu_{2}=3/4$, $\nu_{3}=3/4$,
$\nu_{4}=3/4$, and $\nu_{5}=-5/4$. This result is fully consistent with the
equations (\ref{nucoeff}) which were independently obtained by the
Poincar\'e algebra. Now all of the coefficients of the $H_{S_{1}^2p^2}$ Hamiltonian and of the source terms have been fixed. They read

\begin{eqnarray}
 \alpha_{1}&=&\frac{m_{2}}{4m_{1}^3},\quad\alpha_{2}=0,\quad\alpha_{3}=\frac{3m_{2}}{8m_{1}^3},\quad\alpha_{4}=-\frac{3}{8}\frac{m_{2}}{m_{1}^3},\quad\alpha_{5}=0,\quad\alpha_{6}=-\frac{3m_{2}}{4m_{1}^3},\\
 \beta_{1}&=&\beta_{3}=\beta_{5}=\beta_{6}=0,\quad\beta_{2}=-\frac{3}{4m_{1}m_{2}},\quad\beta_{4}=\frac{9}{4m_{1}m_{2}},\\
 \gamma_{1}&=&\frac{3}{4m_{1}^2},\quad\gamma_{2}=-\frac{9}{4m_{1}^2},\quad\gamma_{3}=0,\quad\gamma_{4}=-\frac{3}{2m_{1}^2},\quad\gamma_{5}=\frac{3}{m_{1}^2},\quad\gamma_{6}=\frac{3}{4m_{1}^2},\quad\gamma_{7}=-\frac{15}{4m_{1}^2},
\end{eqnarray}

\begin{eqnarray}
 \lambda_{1}&=&\frac{1}{4}, \quad\lambda_{2}=\frac{1}{4}, \quad\lambda_{3}=-\frac{1}{24},\quad\lambda_{4}=0,
 \quad\lambda_{5}=-\frac{1}{12},\quad\lambda_{6}=\frac{1}{8},\quad\lambda_{7}=\frac{1}{8},\quad\lambda_{8}=-\frac{1}{12}.
\end{eqnarray}

It may be interesting to mention that in the ADMTT representation the $\lambda_{4}$ source term is not present. Such a term emerges in \cite{BY} with the values
$\mp\frac{1}{12}$ and with $\vct{a}^2$ replaced by the square of
the radius of the throat  of a nonrotating black hole resulting from 
inversion symmetry at the throat. In
Ref. \cite{B} it was shown that a factor $\vct{a}^2$ in the
$\lambda_{4}$ term can also be generated by a specific deviation from
the point-mass structure of a spherically symmetric body.

\section{The test-particle limit}

Next we will consider the test-particle limit of the energy of a
black-hole binary up to the 4th order in $1/r_{12}$. To do this we plug
into Eq. (5.2) of \cite{S85} the static ADM-gauged Kerr metric from
Ref. \cite{HS} labeled by particle number $1$. The mass and momentum
explicitly appearing in this equation will then get the (test-)particle number $2$:

\begin{equation}
 -p_{0}=-\gamma^{ij}_{1}N_{1i}\,p_{2j}+N_{1}\left(m_{2}^2+\gamma^{ij}_{1}p_{2i}p_{2j}\right)^{1/2}\,.
\end{equation}

The inverse 3-metric $\gamma^{ij}$ in ADMTT gauge up to the order $1/r^4$ reads (G=1)

\begin{equation}
 \gamma^{ij}=\left(1-\frac{2m}{r}+\frac{5}{2}\frac{m^2}{r^2}-\frac{5}{2}\frac{m^3}{r^3}+\frac{35}{16}\frac{m^4}{r^4}-\frac{m\vct{a}^2-3m\left(\vct{a}\cdot\vct{n}\right)^2}{r^3}+\frac{7m^2\vct{a}^2}{2r^4}-\frac{9m^2\left(\vct{a}\cdot\vct{n}\right)^2}{r^4}\right)\delta_{ij}-h_{ij}^{TT}\,.
\end{equation}

The result of this operation is

\begin{align}
 \begin{aligned}
  -p_{0}=&m_{2}+\frac{\vct{p}_{2}^2}{2m_{2}}-\frac{m_{1}m_{2}}{r_{12}}-\frac{3}{2}\frac{m_{1}\vct{p}_{2}^2}{m_{2}r_{12}}+\frac{1}{2r_{12}^2}\left(m_{2}m_{1}^2+5\frac{m_{1}^2}{m_{2}}\vct{p}_{2}^2-4m_{1}\vct{a}_{1}\cdot\left(\vct{n}_{12}\times\vct{p}_{2}\right)\right)\\
         &+\frac{1}{r_{12}^3}\Bigg(-\frac{1}{4}m_{2}m_{1}^3-\frac{25}{8}\frac{m_{1}^3}{m_{2}}\vct{p}_{2}^2+\frac{3}{2}m_{1}m_{2}\left(\vct{a}_{1}\cdot\vct{n}_{12}\right)^2+\frac{9}{4}\frac{m_{1}}{m_{2}}\vct{p}_{2}^2\left(\vct{a}_{1}\cdot\vct{n}_{12}\right)^2-\frac{1}{2}m_{1}m_{2}\vct{a}_{1}^2-\frac{3}{4}\frac{m_{1}}{m_{2}}\vct{p}_{2}^2\vct{a}_{1}^2\\
         &\qquad+6m_{1}^2\vct{a}_{1}\cdot\left(\vct{n}_{12}\times\vct{p}_{2}\right)\Bigg)\\
         &+\frac{1}{r_{12}^4}\Bigg(\frac{1}{8}m_{2}m_{1}^4+\frac{105}{32}\frac{m_{1}^4}{m_{2}}\vct{p}_{2}^2-\frac{9}{2}m_{2}m_{1}^2\left(\vct{a}_{1}\cdot\vct{n}_{12}\right)^2+\frac{21}{2}\frac{m_{1}^2}{m_{2}}\left(\vct{p}_{2}\cdot\vct{n}_{12}\right)^2\left(\vct{a}_{1}\cdot\vct{n}_{12}\right)^2-\frac{53}{4}\frac{m_{1}^2}{m_{2}}\vct{p}_{2}^2\left(\vct{a}_{1}\cdot\vct{n}_{12}\right)^2\\
          &\qquad-\frac{7}{4}\frac{m_{1}^2}{m_{2}}\left(\vct{p}_{2}\cdot\vct{a}_{1}\right)^2+\frac{5}{2}m_{2}m_{1}^2\vct{a}_{1}^2-\frac{7}{2}\frac{m_{1}^2}{m_{2}}\left(\vct{p}_{2}\cdot\vct{n}_{12}\right)^2\vct{a}_{1}^2+\frac{23}{4}\frac{m_{1}^2}{m_{2}}\vct{p}_{2}^2\vct{a}_{2}^2+\frac{21}{2}m_{1}^3\vct{a}_{1}\cdot\left(\vct{n}_{12}\times\vct{p}_{2}\right)\\
          &\qquad+5m_{1}\left(\vct{a}_{1}\cdot\vct{n}_{12}\right)^2\vct{a}_{1}\cdot\left(\vct{n}_{12}\times\vct{p}_{2}\right)-m_{1}\vct{a}_{1}^2\vct{a}_{1}\cdot\left(\vct{n}_{12}\times\vct{p}_{2}\right)\Bigg).
 \end{aligned}
\end{align}
Now, from this limit we can also read off the coefficients $\beta_{2}$
and $\beta_{4}$ of Eq. (\ref{p2S2Ham}) in the ADM scheme,
\begin{eqnarray}
 \beta_{2}&=&-\frac{3}{4m_{1}m_{2}}\label{lim1},\\
 \beta_{4}&=&\frac{9}{4m_{1}m_{2}}\label{lim2},
\end{eqnarray}
in confirmation with the previously found ones. This is very strong evidence that we are on the right track with our method.

\section{The 2PN quartic spin Hamiltonian}

To see that there is no $S^4$ Hamiltonian at the order $1/c^4$, we expand
the Kerr metric in ADMTT coordinates up to the orders $1/r^4$ and $a^4$ and show that such terms are not present at all and therefore cannot follow from a source with purely $a^4$ terms. The strategy is the same as in \cite{HS}, but now we allow $a^4$ terms and take the Kerr metric in quasi-isotropic coordinates of Eq. (43) in \cite{HS} up to the aforementioned order and transform to ADMTT coordinates according to the formula

\begin{equation}
 \gamma_{ij}^{ADM}=\gamma_{ij}^{qiso}+\gamma_{ik}^{qiso}\xi^{k}_{,j}+\gamma_{jk}^{qiso}\xi^{k}_{,i}+\gamma_{ij,k}^{qiso}\xi^{k}
\end{equation}
with the 3-metric $\gamma_{ij}^{qiso}$

\begin{eqnarray}
 \gamma_{ij}^{qiso}&=&\gamma_{ij}^{(s)}+\gamma_{ij}^{(2)}+\gamma_{ij}^{(3)}+\gamma_{ij}^{(4)'}\;,\\
 \gamma_{ij}^{(4)'}&=&\gamma_{ij}^{(4)}+\frac{1}{16}\frac{a^4}{r^4}\delta_{ij}\;,
\end{eqnarray}

and the extended transformation vector

\begin{align}
\begin{aligned}
 \xi^{k}=&-\frac{1}{4}\frac{a^2n^{k}}{r}+\frac{1}{2}\frac{(\vct{a}\cdot\vct{n})a^{k}}{r}-\frac{7}{16}\frac{m^2a^2n^{k}}{r^3}+\frac{7}{4}\frac{m^2(\vct{a}\cdot\vct{n})^2n^{k}}{r^3}\\
         &-\frac{1}{16}\frac{a^4 n^{k}}{r^3}+\frac{1}{4}\frac{a^2(\vct{a}\cdot\vct{n})^2n^{k}}{r^3}+\frac{1}{4}\frac{a^2(\vct{a}\cdot\vct{n})a^{k}}{r^3}-\frac{1}{2}\frac{(\vct{a}\cdot\vct{n})^3 a^{k}}{r^3}.
\end{aligned}
\end{align}
We end up with a metric being the same as Eqs. $(50)$ and $(51)$ in
\cite{HS} having all $a^4$ terms transformed away.
This leads to the conclusion

\begin{equation}
 H_{S_1^4} =  H_{S_2^4} = 0
\end{equation}
for the quartic nonlinearities in the spin for the Hamiltonians linear in $G$.

\section{Conclusions}

The Hamiltonians we obtained are summarized as follows:

\begin{align}
 \begin{aligned}
  H_{S_{1}^3p_1}=\frac{G}{r_{12}^4}\Bigg[\vct{S}_{1}\cdot\left(\vct{n}_{12}\times\vct{p}_{1}\right)\left(\frac{m_{2}}{4m_{1}^3}\vct{S}_{1}^2-\frac{5m_{2}}{4m_{1}^3}\left(\vct{S}_{1}\cdot\vct{n}_{12}\right)^2\right)\Bigg],
 \end{aligned}
\end{align}

\begin{align}
 \begin{aligned}
 H_{S_{1}^2S_{2}p_1}=&\frac{G}{r_{12}^4}\bigg[\frac{3}{2m_{1}^2}\vct{S}_{1}^2 \vct{S}_{2}\cdot\left(\vct{n}_{12}\times\vct{p}_{1}\right)+\frac{3}{2m_{1}^2}\left(\vct{S}_{1}\cdot\vct{n}_{12}\right)\vct{S}_{2}\cdot\left(\vct{S}_{1}\times\vct{p}_{1}\right)-\frac{15}{2m_{1}^2}\left(\vct{S}_{1}\cdot\vct{n}_{12}\right)^2\vct{S}_{2}\cdot\left(\vct{n}_{12}\times\vct{p}_{1}\right)\\
&\qquad\quad+\vct{n}_{12}\cdot\left(\vct{S}_{1}\times\vct{S}_{2}\right)\left(\frac{3}{2m_{1}^2}\vct{S}_{1}\cdot\vct{p}_{1}-\frac{15}{2m_{1}^2}\left(\vct{S}_{1}\cdot\vct{n}_{12}\right)\left(\vct{p}_{1}\cdot\vct{n}_{12}\right)\right)\bigg],
 \end{aligned}
\end{align}

\begin{align}
\begin{aligned}
 H_{S_{1}^2p^2}=&\frac{G}{r_{12}^3}\Bigg[\frac{m_{2}}{4m_{1}^3}\left(\vct{p}_{1}\cdot\vct{S}_{1}\right)^2+\frac{3m_{2}}{8m_{1}^3}\left(\vct{p}_{1}\cdot\vct{n}_{12}\right)^{2}\vct{S}_{1}^{2}-\frac{3m_{2}}{8m_{1}^3}\vct{p}_{1}^{2}\left(\vct{S}_{1}\cdot\vct{n}_{12}\right)^2\\
&\qquad-\frac{3m_{2}}{4m_{1}^3}\left(\vct{p}_{1}\cdot\vct{n}_{12}\right)\left(\vct{S}_{1}\cdot\vct{n}_{12}\right)\left(\vct{p}_{1}\cdot\vct{S}_{1}\right)-\frac{3}{4m_{1}m_{2}}\vct{p}_{2}^{2}\vct{S}_{1}^{2}+\frac{9}{4m_{1}m_{2}}\vct{p}_{2}^{2}\left(\vct{S}_{1}\cdot\vct{n}_{12}\right)^2\\
&\qquad +\frac{3}{4m_{1}^2}\left(\vct{p}_{1}\cdot\vct{p}_{2}\right)\vct{S}_{1}^2-\frac{9}{4m_{1}^2}\left(\vct{p}_{1}\cdot\vct{p}_{2}\right)\left(\vct{S}_{1}\cdot\vct{n}_{12}\right)^2-\frac{3}{2m_{1}^2}\left(\vct{p}_{1}\cdot\vct{n}_{12}\right)\left(\vct{p}_{2}\cdot\vct{S}_{1}\right)\left(\vct{S}_{1}\cdot\vct{n}_{12}\right)\\
&\qquad+\frac{3}{m_{1}^2}\left(\vct{p}_{2}\cdot\vct{n}_{12}\right)\left(\vct{p}_{1}\cdot\vct{S}_{1}\right)\left(\vct{S}_{1}\cdot\vct{n}_{12}\right)+\frac{3}{4m_{1}^2}\left(\vct{p}_{1}\cdot\vct{n}_{12}\right)\left(\vct{p}_{2}\cdot\vct{n}_{12}\right)\vct{S}_{1}^2\\
&\qquad-\frac{15}{4m_{1}^2}\left(\vct{p}_{1}\cdot\vct{n}_{12}\right)\left(\vct{p}_{2}\cdot\vct{n}_{12}\right)\left(\vct{S}_{1}\cdot\vct{n}_{12}\right)^2\Bigg],
\end{aligned}
\end{align}
plus the ones with interchanged indices $(1\leftrightarrow2)$, and

\begin{equation}
 H_{S_1^4} =  H_{S_2^4} = 0\,.
\end{equation}

The corresponding sources in the constraint equations take the form

\begin{align}
\begin{aligned}
 \mathcal{H}^{\text{matter}}=&\sum_{b}\Bigg[-\frac{m_{b}}{2}Q_{b}^{ij}\partial_{i}\partial_{j}-\frac{1}{2}\vct{p}_{b}\cdot\left(\vct{a}_{b}\times\vct{\partial}\right)+\left(\gamma^{ij}p_{bi}p_{bj}+m_{b}^2\right)^{1/2}+\frac{1}{8}\frac{\vct{p}_{b}^2}{m_{b}}Q_{b}^{ij}\partial_{i}\partial_{j}\\
                             &\quad+\frac{1}{4m_{b}}(\vct{p}_{b}\cdot\vct{\partial})Q_{b}^{ij}p_{bi}\partial_{j}-\frac{1}{24m_{b}}\vct{a}_{b}^2(\vct{p}_{b}\cdot\vct{\partial})^2+\frac{1}{12}\vct{p}_{b}\cdot\left(\vct{a}_{b}\times\vct{\partial}\right)Q_{b}^{ij}\partial_{i}\partial_{j}\Bigg]\delta_{b},
\end{aligned}
\end{align}

\begin{align}
\begin{aligned}
 \mathcal{H}^{\text{matter}}_{i}=&-2\sum_{b}\Bigg[Q_{b}^{kl}\bigg(-\frac{1}{12}p_{bk}\partial_{l}\partial_{i}+\frac{1}{8}p_{bi}\partial_{k}\partial_{l}+\frac{1}{8}\left(\vct{p}_{b}\cdot\vct{\partial}\right)\delta_{li}\partial_{k}\bigg)\\
&\qquad
+\frac{m_{b}}{4}\left(\vct{a}_{b}\times\vct{\partial}\right)_{i}\left(1-\frac{1}{6}Q_{b}^{kl}\partial_{k}\partial_{l}\right)
-\frac{1}{2}p_{bi}\Bigg]\delta_{b}.
\end{aligned}
\end{align}

The only missing Hamiltonians at the formal order of $1/c^4$ are the
$G^2$ Hamiltonians $H_{S_1^2} + H_{S_2^2}$ resulting
from gravitational nonlinearities of the Einstein field
equations. These Hamiltonians will be presented in a forthcoming paper \cite{SHS}.

\acknowledgments 
The authors thank Jan Steinhoff for helpful discussions. This work is supported by the Deutsche Forschungsgemeinschaft (DFG) through
SFB/TR7 ``Gravitational Wave Astronomy''.

\end{document}